# SYMMETRIC PAIR PLASMAS WITH IMPURITIES AND ACOUSTIC WAVE IN IT

**Barbara Atamaniuk, Krzysztof Żuchowski**
IPPT-PAN Warszawa

A motivation for the study of symmetric pair plasma dynamics lies in the insight they provide into the dynamics of more general plasmas. In the symmetric pair plasma , when temperature both species is the same, acoustic wave are absent. However in presence other species (impurities) in fullerene pair plasmas. for example two-temperature electrons, acoustic waves are possible

## 1.Introduction

Symmetric pair plasmas, consisting of two species with opposite charge and equal masses is an exciting field where not only unexpected phenomena have been experimentally identified, but also where new theoretic problems have been defined. The presence of such physical systems like the early universe. neutron stars, active galactic nuclei, and pulsar magnetosphere makes the understanding of pair plasma dynamics relevant study for astrophysics [1],[2]. Although electron-positron pair plasmas ($e^-, e^+$ plasmas) have since long been in laboratories do not allow life times sufficiently long for the excitation and development of coherent structures like acoustic waves and solitons (the annihilation time is short in comparison to the plasma period). This drawback is not present in the recently available long-lived pair plasmas composed by electron-positron [3] or by single charged fullerene molecules ($C_{60}^-, C_{60}^+$ plasmas) [4],[5]. This poses a unique possibility to investigate the collective behavior of a symmetric pair-ion (mass opposite charged fullerene is almost equal) plasma experimentally under controlled conditions. Electron-hole plasmas ($e^- h^+$ plasmas) in pure semiconductors also are symmetric pair plasmas if effective masses of electrons and holes are equal. Symmetric pair plasmas are the subject of increasing interest among physicists. A motivation for the study of pair plasma dynamics lies in the insight they provide into the dynamics of more general plasmas, often composed of species with very different masses. One of the main difficulties in the numerical research of ordinary electron-ion plasmas is the large difference between the two involved time scales. It was observed in particle-in-cell simulation that a current-carrying Maxwellian plasma composed of two species with not too different masses is unstable for below the threshold predicted by the linear theory. An integration of the full kinetic equation of both species is therefore computationally very expensive and often masked by numerical limitations and simplifications (e.g., modeling one of the species by fluid dynamics) have instead been used to investigate the long-term behavior.

In single and multiion plasmas the large differences between the electron and ion masses typically give rise to different scales, which can be used to disentangle in a natural way long and short wavelength phenomena, or high and low frequency modes. Such distinctions disappear when studying symmetric pair plasma, where the equal masses and opposite charges destroy the scales.

Symmetric pair plasmas demonstrate drastically other collective behavior than ordinary asymmetric electron-ion plasmas ($e^- i^+ plasmas$). A recent theoretical considerations [6] presents a strict proof of impossibility of stationary electrostatic (acoustic) structures such double layers, charged sheaths near electrodes or stationary solitary electrostatic waves in symmetric unmagnitized pair plasmas. The proof is based on the analysis of solutions of momentum and continuity equations of species and Poisson equation with Sagdeev pseudopotential technique. Polytrophic state laws of species with equal coefficients and equal indexes are substituted into the momentum equations. It means that the temperature of species are equal also and thermodynamic equilibrium is assumed. But in [8 ] show how this situation may be changed. In that article main idea is to assume a thermodynamic unequilibrium of pair plasma when temperature of species are not equal ($T_- \neq T_+$). It is proved that in these symmetric pair plasma, but with not equal temperature of species acoustic mode may exist. High temperature laboratory and space plasmas are characterized by the fact that the typical collisional time scale is much longer than the plasma dynamics time scale. As a consequence, these plasmas can be considered in first approximation as collisionlles and the dynamics as Hamiltonian Vlasov-Poisson model since the diffusive scale length is many orders of magnitude smaller than any dynamical or kinetic scale length. In the present study we discuss the properties of collisionless Vlasov-Poisson model in fluid approximation in context of pair plasmas. In symmetric pair plasmas, where both species have the same temperature, acoustic mode are impossible, but in that plasmas exists nonlinear amplitude modulation of electrostatic mode [7]. These mode mixed higher harmonics with basic waves and can gives envelope solitons. It is not a pure acoustic modes according to above remarks. Next we consider acoustic waves in symmetric pair fullerene plasmas ($C_{60}^-, C_{60}^+ plasmas$) with two population of electrons: cold and hot.

## 2. Symmetric pair fullerene plasmas with two populations of electrons

The possibility of symmetric pair fullerene plasmas ($C_{60}^-, C_{60}^+ plasmas$) production involve discussion about existing such plasma completely without of electrons. Therefore it seem interesting to consider fullerene pair plasma with some numbers of electrons with two population: cold and hot. The dynamics of the acoustic waves in a fullerene pair plasmas with two population of electrons system in one-dimensional form is in fluid approximation governed by:

(2.1) $$\frac{\partial N_1}{\partial t} + \frac{\partial}{\partial x}(N_1 U_1) = 0,$$

(2.2) $$\frac{\partial U_1}{\partial t} + U_1 \frac{\partial U_1}{\partial x} = \frac{\partial \phi}{\partial x},$$

(2.3) $$\frac{\partial N_2}{\partial t} + \frac{\partial}{\partial x}(N_2 U_2) = 0,$$

(2.4) $$\frac{\partial U_2}{\partial t} + U_2 \frac{\partial U_2}{\partial x} = -\frac{\partial \phi}{\partial x},$$

(2.5) $$\frac{\partial^2 \phi}{\partial x^2} = N_1 - (1 + \alpha_1 + \alpha_2) N_2 + \alpha_1 e^{\gamma \phi} + \alpha_2 e^{\phi},$$

where $N_1(N_2)$ is the negative (positive) fullerene number density normalized by its equilibrium value $n_{10}(n_{20})$, $U_1(U_2)$ is the negative (positive) fullerene fluid speed normalized by $C = (k_B T_2 / m)^{1/2}$, $\phi$ is the wave potential electrical field normalized by $k_B T_2 / e$, $\gamma = T_2 / T_1$, $m$ is the mass of the fullerene, $T_1(T_2)$ is the temperature of hot (cold) electrons, $k_B$ is the Boltzmann constant, $e$ is the electronic charge, $\alpha_2(\alpha_1)$ electrons cold (hot) number density normalized by $n_{10}$. The time $t$ is normalized by $\omega_{p_1}^{-1} = (m / 4\pi e^2 n_{10})^{1/2}$, and the space variable $x$ is normalized by $\lambda_D = (k_B T_2 / 4\pi e^2 n_{10})^{1/2}$. Let us give a wave-like perturbation to pair plasma and assume that the perturbation propagates with Mach number $M$. The Mach number is the relation between the velocity of wave motion and linear ion sound velocity in unmagnetized plasmas. $M$ is normalized by $C$. Now, we derive the Korteweg-de Vries (K-dV) equation from (2.1)-(2-5) by employing the reductive perturbation technique [9],[10] and the stretched coordinates $\zeta = \varepsilon^{1/2}(x - Mt)$ and $\tau = \varepsilon^{3/2}t$, where $\varepsilon$ is a smallness parameter measuring the weakness of the dispersion.

We can express (2.1)-(2.5) in terms of $\zeta$ and $\tau$ as

(2.6) $\quad\quad\quad \varepsilon^{3/2}\dfrac{\partial N_1}{\partial \tau} - M\varepsilon^{1/2}\dfrac{\partial N_1}{\partial \zeta} + \varepsilon^{1/2}\dfrac{\partial}{\partial \zeta}(N_1 U_1) = 0,$

(2.7) $\quad\quad\quad \varepsilon^{3/2}\dfrac{\partial U_1}{\partial \tau} - M\varepsilon^{1/2}\dfrac{\partial U_1}{\partial \zeta} + \varepsilon^{1/2}U_1\dfrac{\partial U_1}{\partial \zeta} = \varepsilon^{1/2}\dfrac{\partial \phi}{\partial \zeta},$

(2.8) $\quad\quad\quad \varepsilon^{3/2}\dfrac{\partial N_2}{\partial \tau} - M\varepsilon^{1/2}\dfrac{\partial N_2}{\partial \zeta} + \varepsilon^{1/2}\dfrac{\partial}{\partial \zeta}(N_2 U_2) = 0,$

(2.9) $\quad\quad\quad \varepsilon^{3/2}\dfrac{\partial U_2}{\partial \tau} - M\varepsilon^{1/2}\dfrac{\partial U_2}{\partial \zeta} + \varepsilon^{1/2}U_2\dfrac{\partial U_2}{\partial \zeta} = -\varepsilon^{1/2}\dfrac{\partial \phi}{\partial \zeta},$

(2.10) $\quad\quad\quad \varepsilon\dfrac{\partial^2 \phi}{\partial \zeta^2} = N_1 - (1 + \alpha_1 + \alpha_2)N_2 + \alpha_1 e^{\gamma\phi} + \alpha_2 e^{\phi}.$

We can expand the variables $N_1$, $U_1$, $N_2$, $U_2$, and $\phi$ in a power series of $\varepsilon$ as

(2.11) $\quad\quad\quad N_1 = 1 + \varepsilon N_1^{(1)} + \varepsilon^2 N_1^{(2)} + \cdots,$

(2.12) $\quad\quad\quad U_1 = 0 + \varepsilon U_1^{(1)} + \varepsilon^2 U_1^{(2)} + \cdots,$

(2.13) $\quad\quad\quad N_2 = 1 + \varepsilon N_2^{(1)} + \varepsilon^2 N_2^{(2)} \cdots,$

(2.14) $\quad\quad\quad U_2 = 0 + \varepsilon U_2^{(1)} + \varepsilon^2 U_2^{(2)} + \cdots,$

(2.15) $\quad\quad\quad \phi = 0 + \varepsilon \phi^{(1)} + \varepsilon^2 \phi^{(2)} + \cdots.$

Electrons number density is treated as a parameters.

Now, using (2.11)-(2.15) in (2.6)-(2.10) and taking the coefficient of $\varepsilon^{3/2}$ from (2.6)-(2.9) and $\varepsilon$ from (2.10) we get

(2.16) $\quad\quad\quad N_1^{(1)} = U_1^{(1)}/M,$

(2.17) $\quad\quad\quad U_1^{(1)} = -\phi^{(1)}/M,$

(2.18) $\quad\quad\quad N_2^{(1)} = U_2^{(1)}/M,$

(2.19) $\quad\quad\quad U_2^{(1)} = \phi^{(1)}/M,$

(2.20) $\quad\quad\quad N_1^{(1)} - (\alpha_1 + \alpha_2)N_2^{(1)} + \alpha_1\gamma\phi^{(1)} + \alpha_2\phi^{(2)} = 0.$

Now, using (2.16)-(2.20), we have

(2.21) $\quad\quad\quad N_1^{(1)} = -\phi^{(1)}/M^2,$

(2.22) $\quad\quad\quad N_2^{(2)} = \phi^{(1)}/M^2,$

(2.23) $\quad\quad\quad M^2 = \dfrac{2 + \alpha_1 + \alpha_2}{\gamma\alpha_1 + \alpha_2}.$

Equation (2.23) is the linear dispersion relation for acoustic waves propagating in our fullerene pair plasma model. Now substituting (2.11)-(2.15) into (2.6)-(2.10) and equating the coefficient of $\varepsilon^{3/2}$ from (2.6)-(2.9) and $\varepsilon^2$ from (2.10), one obtains

(2.24) $\quad\quad\quad \dfrac{\partial N_1^{(1)}}{\partial \tau} - M\dfrac{\partial N_1^{(2)}}{\partial \zeta} + \dfrac{\partial U_1^{(2)}}{\partial \zeta} + \dfrac{\partial}{\partial \zeta}\left[N_1^{(1)} U_1^{(1)}\right] = 0,$

(2.25) $\quad\quad\quad \dfrac{\partial U_1^{(1)}}{\partial \tau} - M\dfrac{\partial U_1^{(2)}}{\partial \zeta} + U_1^{(1)}\dfrac{\partial U_1^{(1)}}{\partial \zeta} = \dfrac{\partial \phi^{(2)}}{\partial \zeta},$

(2.26) $$\frac{\partial N_2^{(1)}}{\partial \tau} - M \frac{\partial N_2^{(2)}}{\partial \zeta} + \frac{\partial U_2^{(2)}}{\partial \zeta} + \frac{\partial}{\partial \zeta}\left[N_2^{(1)}U_2^{(1)}\right] = 0,$$

(2.27) $$\frac{\partial U_2^{(1)}}{\partial \tau} - M \frac{\partial U_2^{(2)}}{\partial \zeta} + U_2^{(1)}\frac{\partial U_2^{(1)}}{\partial \zeta} = -\frac{\partial \phi^{(2)}}{\partial \zeta},$$

(2.28) $$\frac{\partial^2 \phi^{(1)}}{\partial \zeta^2} = N_1^{(2)} - (1+\alpha_1+\alpha_2)N_2^{(2)} + \alpha_1\gamma\phi^{(2)} + \frac{1}{2}\alpha_1\gamma^2\left[\phi^{(1)}\right]^2 + \alpha_2\phi^{(2)} - \frac{1}{2}\alpha_2\left[\phi^{(1)}\right]^2.$$

Now, using above equation, and eliminating $N_1^{(2)}, N_2^{(2)}, U_1^{(2)}, U_2^{(2)}$, and $\phi^{(2)}$, we finally obtain

(2.29) $$\frac{\partial \phi^{(1)}}{\partial \tau} + A\phi^{(1)}\frac{\partial \phi^{(1)}}{\partial \zeta} + B\frac{\partial^3 \phi^{(1)}}{\partial \zeta^3} = 0,$$

Where the nonlinear coefficient $A$ and the dispersion coefficient $B$ are given by

(2.30) $$A = \frac{1}{2M[2+\alpha_1+\alpha_2]}\left[3(1+\alpha_1+\alpha_2) - 3 - M^4\left(\gamma^2\alpha_1+\alpha_2\right)\right],$$

(2.31) $$B = \frac{M^3}{2(2+\alpha_1+\alpha_2)}.$$

Equation (2.29) is the K-dV equation describing the nonlinear propagation of the acoustic waves in our fullerene pair plasmas with electron impurities. The steady state solution of this K-dV equation may be obtained by transforming the independent variables $\zeta$ and $\tau$ to $\xi = \zeta - M_0\tau$ and $\tau = \tau$ where $M_0$ is a constant velocity normalized by $C$, and imposing the appropriate boundary conditions: $\phi^{(1)} \to 0, \partial\phi^{(1)}/\partial\xi \to 0, \partial^2\phi^{(1)}/\partial\xi^2 \to 0$ at $\xi \to \pm\infty$. Thus, the steady state solution of the K-dV equation is

(2.32) $$\phi^{(1)} = \phi_m \operatorname{sech}^2(\xi/\Delta),$$

Where the amplitude $\phi_m$ which is normalized by $k_B T_2/e$ and the width $\Delta$ normalized by $\lambda_D$ are given by

(2.33) $\phi_m = 3M_0/A,$

(2.34) $\Delta = \sqrt{4B/M_0}.$

It is clear from (2.32)-(2.34) that as $M_0$ increases, the amplitude (width) of the solitary waves increases (decreases). It is evident from (2.30), (2.32), and (2.33) that the solitary potential profile is positive (negative) if $A > 0$ ($A < 0$). If a number density of electrons $\alpha_1$ and $\alpha_2$ decreases ($\alpha_1 \Box 1, \alpha_2 \Box 1$) from (2.30), (2.31),(2.33) and (2.34) we have that $\Delta$ increases and $\phi_m$ decreases which denote that the soliton wave disappear (decays). In accordance that in a pure symmetric pair plasma the acoustic structures are absent.